\documentclass[aps,prl, twocolumn,showpacs,floats,superscriptaddress]{revtex4}
\usepackage{epsfig,color}

\begin{document}

\title{Criticality in Trapped Atomic Systems}

\author{L. Pollet}
\affiliation{Physics Department, Harvard University, Cambridge MA 02138, USA}
\author{N.V. Prokof'ev}
\affiliation{Department of Physics, University of Massachusetts,
Amherst, MA 01003, USA}
\affiliation{Russian Research Center ``Kurchatov Institute'',
123182 Moscow, Russia}

\author{B.V. Svistunov}
\affiliation{Department of Physics, University of Massachusetts,
Amherst, MA 01003, USA}
\affiliation{Russian Research Center ``Kurchatov Institute'',
123182 Moscow, Russia}

\date{\today}

\begin{abstract}
We discuss generic limits posed by the trap in atomic systems on the accurate determination of critical parameters for second-order phase
transitions, from which we deduce optimal protocols to extract them.
We show that under current experimental conditions the in-situ density profiles are barely suitable for an accurate study of
critical points in the strongly correlated regime. Contrary to recent claims, the proper analysis of time-of-fight images yields critical parameters accurately.
\end{abstract}

\pacs{64.70.Tg, 03.67.Ac, 67.85.Hj, 74.20.-z}
\maketitle

Recent breakthroughs in cold atom lattice experiments have paved the road to reliable studies of 
condensed matter physics by emulating its intractable models. Although the identification of phases in present experiments is nearly universally accepted, the mesoscopic size and the presence of a curved potential have long plagued our understanding of criticality in optical lattice experiments. This has sparked vivid debates~\cite{Diener07, Gerbier07, Kato08, Zhou09, Trotzky09} for even the simple normal-to-superfluid (N-SF) U(1) symmetry breaking transition for interacting bosons. 
In order to be reliable quantum emulators, cold atom experiments need a generic protocol 
for studying critical behavior, and they need to do so without invoking numerical simulations.

The original experiments in optical lattices \cite{Greiner02} were based on time-of-flight
(TOF) images and the emergence of a `sharp' peak in the momentum distribution.
The approach is adequate for identifying superfluid and normal phases,
but it lacks an explicit procedure for locating the phase boundary, since peaks already develop in the normal phase due to an increasing
correlation length \cite{Kashurnikov02, Wessel04, Pollet04, Schroll04}. It  has been proposed in Ref.~\cite{Kollath04} that one should trace the {\it evolution of the peak shape}, and use the increase in the peak width as the transition signature. The peak width keeps decreasing across the transition point however, and this leaves room for subjective data analysis and thus increased error bars.
Monitoring both the peak width and the number of particles in the low-momentum peak 
has been used recently in measuring the N-SF transition temperature and 
its suppression on approach to the Mott insulating phase \cite{Trotzky09}.

Difficulties with the analysis of TOF images and advances in single-site resolution detection methods~\cite{Gemelke09, Bakr09} prompted the authors of Ref.~\cite{Zhou09} to propose that finite-temperature critical points of strongly correlated quantum models emulated by an optical
lattice experiment can be generically deduced from cusps in the derivative of the 
density profile of atoms in the trap with respect
to the external potential, $\kappa (r) = -dn(r)/dV(r)$. Within the local density 
approximation (LDA) approach this quantity coincides with the compressibility $dn/d\mu $.
However, the proposal of Ref.~\cite{Zhou09} was shown not to work under realistic 
experimental conditions \cite{Comment}.

The goal of this Letter is to answer qualitatively and quantitatively whether 
determining critical parameters in a trapped system using available experimental techniques 
is feasible. First, we employ the theory of finite-size effects in critical phenomena 
discussed recently for trapped systems by Campostrini and Vicari \cite{ettore} 
to set theoretical limits on the accurate location of phase transition points 
in a given trap for a generic,
strongly correlated system. Second, we demonstrate that LDA violations in density 
profiles indicate critical behavior, not LDA itself. The calculations show, 
however, that experimental data should have extremely small error bars, 
well below what is currently available. Taking the numerical derivative of the 
density profiles is of no help, and is always either too noisy or lacking 
features associated with criticality.
Third, we demonstrate that detection methods coupling directly to critical modes, 
{\it e.g.}, TOF images in the case of the N-SF transition, do allow one to reach the
theoretical limit. From the analysis of the central peak {\it shape} 
width we construct a quantity with a sharp minimum at the critical point. 
Our considerations apply to any scale-invariant second-order phase transition.

We simulate the 3D Bose-Hubbard model,
\begin{equation}
H = - t \sum_{<ij>} b_{i}^{\dag} b_{j}^{\;} +\frac{U}{2}
\sum_{i}\: n_{i}(n_{i}-1) - \sum_{i}\:(\mu - V_i) n_{i} \; ,
\label{BH}
\end{equation}
where $t$ is the matrix element describing the hopping of bosons between nearest-neighbor sites, $U$ is the on-site repulsion, $V_i$ is the trapping potential, and  $n_{i}=b_{i}^{\dag} b_{i}^{\;}$ is the on-site occupation number in terms of bosonic creation and annihilation operators. In what follows, we use $t$ as the unit of energy.

Let the phase transition occur around the chemical potential
$\mu (r_c) =\mu_c$, where $\mu_c$ is the critical point
in a homogeneous system, and $r_c$ is situated
away from the trap center. By the very nature of second-order phase transitions, characterized by a divergent correlation radius $\xi$ at $\mu_c$,  there exists a finite shell
centered at $r=r_c$ in which LDA fails. Indeed, LDA  implies  quasi-homogeneity of the system
when the change of thermodynamic properties of  the system is negligible at the
distance of the order of correlation radius. Clearly, the quasi-homogeneity is preserved as long as
$|\xi(r \pm \xi(r)) - \xi(r)| \, \ll \, \xi(r) $,
and is violated otherwise, the latter inevitably happening when $|r-r_c| \lesssim  \xi(r)$,
where we have used the shorthand notation $\xi(r)\equiv \xi(\mu(r))$. The condition
\begin{equation}
|r_*-r_c| \, \sim\, \xi(r_*)\, \equiv\, \xi_*
\label{r_star}
\end{equation}
defines the position of the boundaries $r_*$ of the LDA violated region,  and
the largest possible correlation radius.
The value of $\xi_*$ controls the rounding effects, similar to a finite-size system, leading to an
intrinsic uncertainty in determining critical parameters: $\delta \mu_c \propto \xi_*^{-1/\nu}$, where $\nu$ is the
correlation length exponent.  To cast all relations in dimensionless
form we introduce a
typical scale for the chemical potential, $\mu_0$, which we choose such that when $\mu$ is
changed by $\mu_0$ the system density changes by
a factor of two. We also introduce the healing length $d$ as the
typical length scale. Then,  $(\delta \mu_c  / \mu_0 ) \sim (d/\xi_*)^{1/\nu}  $.

In a constant
potential gradient, Eq.~(\ref{r_star}) implies
\begin{equation}
\xi_* \, =\, d  \left( {\mu_0 \over \xi_*  \nabla \mu } \right)^{\nu} \, ~\longrightarrow  ~
{\xi_* \over d} \sim \left( \frac{\mu_0}{d \nabla \mu }\right)^{\nu/(1+\nu)} \, .
\label{eq1}
\end{equation}
Assuming spherically symmetric  harmonic
confinement, we have $\nabla \mu = m\omega^2 r_c$.  By introducing the typical
cloud size using $m\omega^2 R^2 = \mu_0$, the final estimate for $\delta \mu_c$
can be identically rewritten as
\begin{equation}
\frac{\delta \mu_c}{ \mu_0} \sim \left( \frac{dr_c}{R^2}\right)^{1/(1+\nu)}
\;.
\label{eq3}
\end{equation}

Obviously, the best accuracy  is found when the critical point
is located at the trap center where the chemical potential gradient is zero.
In this case the derivation has to be repeated along the same lines using
$\delta \mu_c \sim m\omega^2 \xi_*^2$ for the change of the chemical
potential over the correlation length. The self-consistent solution then 
takes the form
\begin{equation}
\frac{\xi_*}{d} \sim \left( \frac{R}{d}\right)^{2\nu/(2+\nu)}
\;, \;\;\;\;\;
\frac{\delta \mu_c}{ \mu_0} \sim \left( \frac{d}{R}\right)^{2/(2\nu +1)} \;,
\label{eq4}
\end{equation}
reproducing the result of Ref. \cite{ettore} for scaling of the correlation 
length with the trap size.
For the N-SF transition ($\nu=0.6715$) in the strongly correlated Bose
lattice system, the size is $R/d \sim 100$, which leads to the conclusion
that the critical point may, in principle, be determined with a relative
accuracy of a  few percent. Note that this estimate is assuming that system
properties are known exactly, {\it i.e.}, it does {\it not} take into account
experimental noise which we consider next.

Density profiles $n(r)$ can be measured with single-site resolution~\cite{Gemelke09, Bakr09}, and,
under ideal experimental conditions, further averaged over about a hundred
shots. Unfortunately, the density and its derivative are continuous functions
across the N-SF transition point. More precisely, the critical contribution
to compressibility, $\kappa = dn/d\mu $, is of the form
$A (\vert \mu - \mu_c \vert /\mu_0 )^{\alpha }$, where $\alpha = d\nu -2=0.015$
with slightly different amplitudes $A$ on the two sides of the transition.
For the $U/t=24$, $T/t=2.4$ example considered below the value of $A$
is about $0.5/t$. To understand the effect of critical fluctuations
on density profiles we invert the definition of compressibility and
calculate the deviation of the density at point $r_c$ from the
thermodynamic limit value $n_c$ as
\begin{equation}
n(r_c) - n_c \sim -A \int_0^{\delta \mu_c} dx
\left[ \left( \frac{x}{\mu_0}\right)^{\alpha } -
\left( \frac{\delta \mu_c}{\mu_0}\right)^{\alpha } \right]
\;.
\label{eq5}
\end{equation}
This equation accounts for  LDA violations
due to the difference between the thermodynamic expression for the critical
part of compressibility and its maximal value in a system of size $\xi_*$.
Given the small value of $\alpha$ for most continuous phase transitions we can write
\begin{equation}
n(r_c) = n_c + \frac{\alpha A \mu_0}{1+\alpha} \left(\frac{\delta \mu_c}{\mu_0}\right)^{1+\alpha } \approx n_c +\alpha A \delta \mu_c
\;,
\label{eq6}
\end{equation}
which allows us to estimate the size of the effect in realistic traps
using Eqs.~(\ref{eq3}) and (\ref{eq4}). The largest LDA violation
is expected at the trap perimeter where the chemical
potential gradients are large.

\begin{figure}[htb]
\includegraphics[width=0.8\columnwidth]{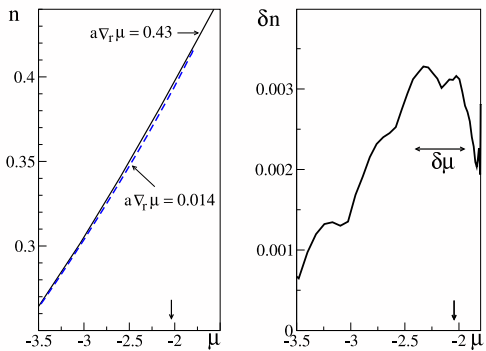}
\caption{(Color online). Density profiles in the vicinity of the critical point
(left panel) demonstrating violations of the LDA approximation
in the presence of finite chemical potential gradient.
In the right panel we plot the difference between the two curves.
The critical point $\mu_c =-2.037$ is marked by an arrow. } \label{fig1}
\end{figure}

Since the critical value of the density is not known {\it a priori},
one should measure density profiles
with different global chemical
potentials $\mu_i(r=0)$, or in traps with different confinement
frequencies $\omega_i$, convert them to $n(\mu, i)$ curves,
and plot them in the same figure. The expectation is that densities measured
at different chemical potential gradients will show the largest deviation
from each other in the critical region. Unfortunately, the size of the effect is
supposed to be rather small in the strongly correlated regime where
$A\mu_0$ is not large, see Eq.~(\ref{eq6}).
The best possible contrast can be found when the profile $n(\mu, 2)$ with the critical region in the center
is subtracted from the profile with the critical region in the perimeter $n(\mu, 1)$, yielding $\delta n (\mu) = n(\mu, 1) - n(\mu, 2)$,
and thus using the system with the smallest size of the critical region as the reference curve.
From Eqs.~(\ref{eq3}), (\ref{eq4}), and (\ref{eq6}) we deduce that
$\delta n (\mu_c)$ obtained for profiles with the critical point
at the trap perimeter and the trap center is about
\begin{equation}
\delta n(\mu_c) \sim  \alpha A \mu_0 \left[\left( \frac{d}{R}\right)^{1/(1+\nu)} - \left( \frac{d}{R}\right)^{2/(2\nu +1)} \right]
\;,
\label{eq7}
\end{equation}
which for the N-SF transition with size $R/d=100$
is as small as $ 4\times 10^{-4} A \mu_0$, {\it i.e.,} it is about a few tenths
of a percent only! In Fig.~\ref{fig1} we show the result of the Monte Carlo
simulation \cite{Prokofev98, Pollet07} of a trapped system for $U/t=24$,
$T/t=2.4$, and $m\omega^2 a^2/2t =0.001$, where $a$ is the lattice constant.
As expected, the largest LDA violations are observed for the curve
with the critical point closest to the trap center. From the position
of the maximum we deduce that $\mu_c=-2.3 \pm 0.4$, and
$n_c = 0.37 \pm 0.05$.

It is thus possible, in principle, to determine critical points with
an accuracy of about
10-20\% by examining LDA violations in density profiles provided they are
measured with extremely high accuracy at the trap center.
[Note that we implicitly assume that experiments are able to
determine the system temperature and chemical potential
by other methods, {\it e.g.,} using the equation of state at the trap perimeter,
entropy matching, etc.]
It is evident  from Fig.~\ref{fig1} that density profiles have to be
measured with a relative accuracy of $0.001$.
As far as we know such accuracy is not available in current experiments.
The noise in the density at the trap
center is determined by the number of experimental runs because
in one measurement $\delta n_i \sim n_i \sim 1$. Optimistically,
the error bars on radial density are about $\sigma (0) \sim 0.1$
at the trap center, quickly reducing to $\sigma (r) \sim \sigma (0)d/4r$
at a distance $r$ due to radial averaging. Hypothetically, it would be
possible to have data in three-dimensional systems
with accuracy of $0.001$ at a distance $r/d>25$ with signal-to-noise
ration close to unity if all other experimental uncertainties are eliminated.

\begin{figure}[htb]
\includegraphics[width=0.8\columnwidth]{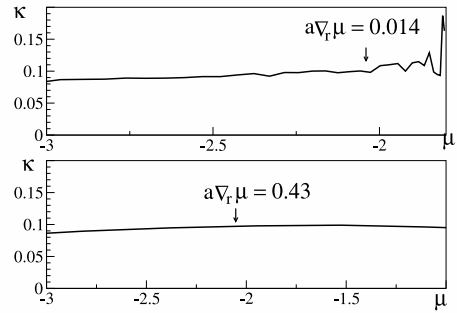}
\caption{(Color online). Local compressibility  across the critical point
at $\mu_c =-2.037$ (marked with an arrow). In the upper panel the
critical region is close to the trap center. In the lower
panel the critical region is close to the trap perimeter and is subject to
a larger chemical potential gradient. Compressibility data
were deduced from curves shown in Fig.~\ref{fig1}. }
\label{fig2}
\end{figure}

\begin{figure}[htb]
\includegraphics[angle=-90, width=0.8\columnwidth]{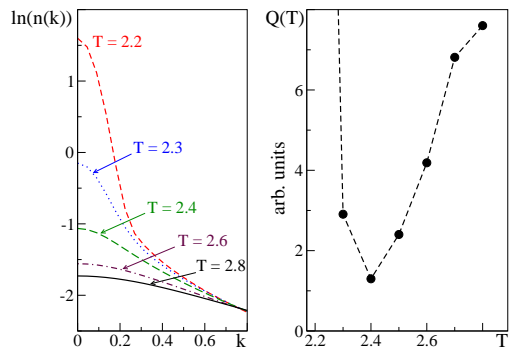}
\caption{(Color online). Momentum distributions and the analysis of the low-momentum peak shape as  described in the text. Here,  $s=2$.
} \label{fig3}
\end{figure}

Our final note concerns density profile derivatives
$\kappa = -dn/dV(r)$. It is clear from the beginning that measures based on
integrals of $\kappa$, {\it i.e.,} the LDA violations discussed above, contain all the information and are more
sensitive because they efficiently average out the noise in numerical derivatives.
Indeed, in the upper panel of Fig.~\ref{fig2} we show
that $\kappa (r)$ deduced from the profile giving the
largest signal in Fig.~\ref{fig1} is meaningless.
Derivatives taken away from the trap center are less noisy, but they also lack any
signature of the critical point due to the rounding of critical singularities by
large chemical potential gradients, see the lower panel in Fig.~\ref{fig2}.
In general, LDA violations are a direct consequence of a diverging length scale
while rounded peaks in derivatives are not necessarily originating from
criticality and may thus be misleading, especially when searching for novel phases.

There is thus no viable alternative to directly studying critical
modes. For the N-SF transition they are encoded in the low-momentum part
of the distribution $n(k)$ measured in TOF images.
Here, we discuss the best case for the accuracy when the
phase transition starts in the middle of the trap. In a homogeneous
system, on approach to the critical
point from the normal phase
the peak in the momentum distribution around $k=0$ is described by
\begin{equation}
n(k)\sim \left\{
\begin{array}{ll}
 k^{\eta -2} & \;\; \mbox{ for }   \;\; \xi ^{-1} \ll k \ll a^{-1} \\
 \xi ^{2- \eta} & \;\; \mbox{ for } \;\; k \ll \xi ^{-1}
\end{array}
\right.
\;,
\label{eq8}
\end{equation}
where $\eta $ is the correlation function exponent (for most
critical points $\eta$ is positive and small; for the N-SF
transition $\eta = 0.038$). In a trap, we have to account
for the fact that $\xi $ depends on the position in space
since for a given detuning at the trap center, $\Delta \mu (T)$,
we have
$\xi (r) \sim [\Delta \mu (T)+ m\omega^2r^2/2 ]^{-\nu }$
at some distance $r$. Although shells away from the center
have smaller $\xi$ (in the normal phase) they have an increasingly larger volume.
Curiously, in a wide trap the peak
amplitude, $P=n(k=0)$, at the critical point
is determined by regions with short correlation
length because the LDA integral for $P$
\begin{equation}
P \sim \int [\xi (r)]^{2-\eta} d^3 r \propto
\int  \frac{d^3r}{ (r^2)^{(2-\eta)\nu} }
\, ,
\label{eq9}
\end{equation}
diverges at large $r$ meaning that the major contribution comes from
a region with $\xi \sim d$, {\it a posteriori} justifying
the use of LDA for evaluating the peak amplitude.

Thus, at $T>T_c$ the TOF image remains relatively broad and the critical
region at the trap center is {\it not} dominating in the signal
amplitude in the theoretical limit of a very broad trap. [ As we show in Fig.~\ref{fig3},
it remains a sizable $50\%$  of the total signal for a realistic trap with $R/d \sim 100$ ].
Past the critical point, the low-momentum part of the distribution
develops the condensate peak smeared out by the presence of the trapping
potential and the finite size of the SF domain superimposed on the
broader peak representing the critical shell separating normal
and superfluid phases.

The all-decisive question is how to ``read out'' the critical point.
Our proposal is
based on monitoring the development of the bimodal structure
at low momenta across the transition point.
We note that the {\it shape} (in contrast to the value of the signal) of $n(k)$ at small momenta is
determined by the critical region at the center because
the first (and also higher-order) derivative, $dn(k)/dk$,
is described by an integral over $d^3r$ which diverges
at the lower limit! This suggests the following protocol
for analyzing the data: \\
(i)  find the momentum $k_{\rm max}(T)$ at which the absolute value of the first derivative $dn(k)/dk$
has a maximum; \\
(ii) construct the ``amplitude of the critical signal'' as the difference
 $P_c(T)=n(k=0)-n(k_{\rm max})$;   \\
(iii) plot $Q(T)=P_c (T) k_{\rm max}^s (T)$, with some exponent $s>2-\eta $,
as a function of temperature to ``read out'' the critical point from the
minimum in $Q(T)$.

By construction, $P_c \propto \xi^{2-\eta }$ and $k_{\rm max} \propto 1/\xi $,
and thus the choice of $s$ is ensuring that $Q(T)$ is decreasing
on the normal side of the transition as $T$ approaches $T_c$.
On the superfluid side, the formation of the  condensate at the trap center results in a sharp increase of the peak amplitude $n(k=0)$ and thus $Q(T)$.
The critical point is then estimated from the position of the minimum in $Q(T)$.
In Fig.~\ref{fig3} we present results of Monte Carlo
simulations for our reference system, $U/t=24$,
done at various temperatures in a trap with $\mu (r=0) = -2.037$, and
$m\omega^2 a^2/2 = 0.0033 t$. The original data for $n(k)$ shown in the
left panel were analyzed to produce the $Q(T)$ curve shown in the right panel.
The position of the minimum at $T/t=2.4$ gives the correct
value of the critical temperature with an accuracy of $\sim 5 \%$ limited
by the temperature grid; otherwise the error bars are obtained from
the FWHM of $1/Q$.

If the momentum resolution of TOF images is better
than the limit $\pi /\xi_*$ set by Eq.~(\ref{eq4}), including
finite time-of-flight \cite{Gerbier08},
optical broadening and resolution effects~\cite{Trotzky09},
then we believe the same accuracy would be possible experimentally.
For columnar images, the exponent $s$ can be reduced to $> 1-\eta$.

In conclusion, critical points in strongly correlated states of
trapped atomic systems can be measured with relative
accuracy of a few percent. The most sensitive signal-to-noise probes
are based on critical modes with diverging correlation functions.

This work was supported by the Swiss National Science Foundation,  the National Science Foundation under Grant PHY-0653183, and a grant from the Army Research Office with funding from the DARPA OLE program.
Simulations were performed on the CM cluster at UMass, Amherst, and
Brutus cluster at ETH Zurich; use was made of the ALPS libraries for the error evaluation~\cite{ALPS}.

\end{document}